\documentclass[conference]{IEEEtran}
\ifCLASSINFOpdf
\else
\fi
\usepackage{cite,url}
 \usepackage [pdftex]{graphicx}
  \graphicspath{ {./figures/} }
\usepackage{algorithmic,algorithm2e}
\usepackage{amsmath,amssymb,amsfonts}
\usepackage{ifthen}
\usepackage{textcomp}
\usepackage{multirow}
\usepackage{gensymb}
\usepackage{caption}
\def\BibTeX{{\rm B\kern-.05em{\sc i\kern-.025em b}\kern-.08em
    T\kern-.1667em\lower.7ex\hbox{E}\kern-.125emX}}
\DeclareGraphicsExtensions{.pdf,.jpeg,.png}

\begin{document}
%
\title{Diabetic Retinopathy Grading System Based on Transfer Learning}

\author{\IEEEauthorblockN{Eman AbdelMaksoud}
\IEEEauthorblockA{Information Systems Dept.\\
Faculty of Computers and Information\\
 Mansoura University, Mansoura\\
Email: eng.eman.te@gmail.com}
\and

\IEEEauthorblockN{Sherif Barakat}
\IEEEauthorblockA{Information Systems Dept.\\
Faculty of Computers and Information\\
Mansoura University, Mansoura\\
Email: sherief.barakat@mans.edu.eg}
\and

\IEEEauthorblockN{Mohammed Elmogy}
\IEEEauthorblockA{Information Technology Dept.\\
Faculty of Computers and Information\\
Mansoura University, Mansoura, P.O.35516\\
e-mail:melmogy@mans.edu.eg}}


%


\maketitle

%

\begin{abstract}
Much effort is being made by the researchers in order to detect and diagnose diabetic retinopathy (DR) accurately automatically. The disease is very dangerous as it can cause blindness suddenly if it is not continuously screened. Therefore, many computers aided diagnosis (CAD) systems have been developed to diagnose the various DR grades. Recently, many CAD systems based on deep learning (DL) methods have been adopted to get deep learning merits in diagnosing the pathological abnormalities of DR disease. In this paper, we present a full based-DL CAD system depending on multi-label classification. In the proposed DL CAD system, we present a customized efficientNet model in order to diagnose the early and advanced grades of the DR disease. Learning transfer is very useful in training small datasets. We utilized IDRiD dataset. It is a multi-label dataset. The experiments manifest that the proposed DL CAD system is robust, reliable, and deigns promising results in detecting and grading DR. The proposed system achieved accuracy (ACC) equals 86\%, and the Dice similarity coefficient (DSC) equals 78.45\%.
\end{abstract}

%
\maketitle
\section{Introduction}
\begin{figure*}[ht]
    \centering
\includegraphics[width=\textwidth]{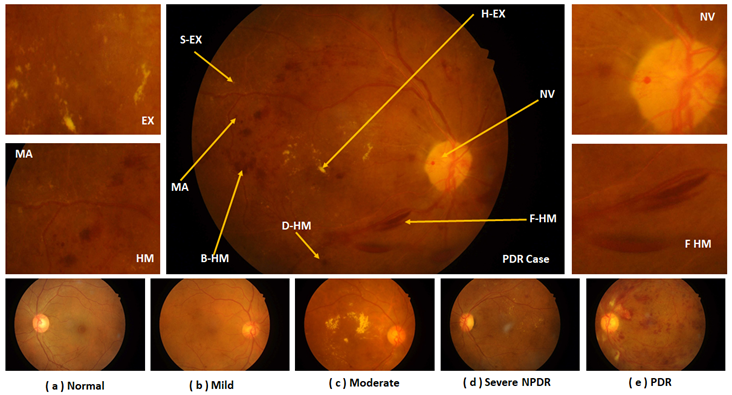}
    \caption{The different DR signs of the color fundus image (EX, MA, HM, and NV), (a) normal case, (b) the mild case, (c) moderate case, (d) severe case, and (e) PDR case.}
    \label{fig:examples}
    \end{figure*}

The retina is an essential part of the human eye. It covers the back of the eye. It detects the lights and sends signals to the brain through the optic nerve (ON). The retina is susceptible to damage due to excess blood sugar caused by diabetes. Diabetes damages the blood vessels (BV) all over the body. The retina's damage starts when the sugar prohibits the tiny BV that goes to it, causing them to leak fluid or bleed. In this case, the eye begins to grow new weak BV that can leak or bleed easily as they do not work well. The caused damage of the retina is called diabetic retinopathy (DR).  Anyone with diabetes can get DR, even with gestational diabetes that can develop during pregnancy \cite{1}. DR, in its advanced stage, can cause blindness. Therefore, the patient with diabetes should constantly control the blood, blood pressure, and cholesterol levels by making continuous checking up and auditing.
The leakage and bleeding causes are appearing some DR lesions with different features on the retina and BVs such as micro and macro aneurysms (MA), dot, blot, and flame hemorrhages (HM), exudates (EX), cotton wool spots (CWS), venous beading (VB) and neovascularization (NV). These lesions specify the DR grades categorized as early or mild, moderate, advanced, or severe. All of these grades are found in a general category called non-proliferative DR (NPDR). In this category, the treatment is possible. On the other hand, the severe grade on NPDR is an indicator of the difficult DR category that is proliferative DR (PDR). When the patient reaches this category, the vision may be lost suddenly\cite{2}. 
In NPDR is determined by appearing MA, dot HM, scattered EX, CWS, and VB.  On the other side, PDR is determined by appearing NV at the optic disc (OD) called NVDs or elsewhere called NVEs, large blot HM, and VB. MA occurs on the superficial of the retina as small red points. EX is caused by leakage of the plasma. It looks like bright yellowish spots with sharp margins. It can occur on the retina outer layer. HM looks like the MA if it is a small dot or blot, HM, and it is in the deepest retina layer. The large flame HM is also called splinter HM, is found in the superficial of the retina layer. CWS is small white with pale and irregular margins. It is caused by nerve fiber swelling\cite{3}. Among the eye imaging modalities, we utilized the most famous one, the fundus modality. It captures the anatomic eye structure helping detect and diagnose DR. Fundus photography is useful in DR screening. It documents the retinal disease timely. It includes different types, standard, wide-field, and stereoscopic. Besides, it is cheaper and useful for monitoring the DR progression over time \cite{4,5}. Fig. \ref{fig:examples} shows the different DR signs of the retinal color fundus images and the four DR grades (mild, moderate, severe NPDR, and PDR), where the F-HM is the flame HM, D-HM is the dot HM, H-EX is hard EX, and S-EX is soft EX.


Regular retinal screening for diabetes patients is still difficult, especially in poor developing countries. Retinal photography with grading and interpretation by ophthalmologists and trained graders is widely important for DR screening. However, the availability of retina experts and well-trained graders is a significant limitation. Even when they are available, there could be a time delay in graders submitting their DR grading and advice due to their busy schedule. This leads to delayed interpretation and loss to follow-up, miscommunication, and delay in proceeding for DR severity management.
Because the detection and grading of DR disease are crucial, consumes much time, and burdens the ophthalmologists, many systems have been developed to detect the DR grades. These systems may entirely base on deep learning (DL) such as\cite{3,6,7} or based on conventional methods such as\cite{8} or make the only classification by DL, such as\cite{9}. Recently, various DL architectures have been developed, such as Restricted Boltzmann Machines (RBMs), deep neural network (DNN), and Convolutional Neural Networks (CNNs). CNN has achieved great success in many applications, such as medical image analysis and multi-label (ML) classification. That is because CNN can extract features more discriminative for large-scale datasets. 
The main reason for utilizing the DL models is that the various feature in the DR fundus images. It is challenging to extract all abnormalities in the fundus images such as BV, H-EX, S-EX, B-HM, F-HM, D-HM, NV, CWS, VB, and others. Moreover, there are many signs, as shown in Fig\ref{fig:examples}, take the same features as each other. Small MAs are like D-HM, large MA is like B-HM in color and shape. On the other hand, H-EXs and OD are the same in the intensity and shape features. HM, MA, and BV take the same color feature. Sometimes, CWS and EXs are the same in the color feature. According to the similarity of the features mentioned above, the traditional (hand-crafted) feature extraction methods such as scale-invariant feature transform (SIFT), histogram oriented gradients (HOG), and bag-of-words histograms (BoW) are unsuitable for fundus images.
The fundus images are with small differences, as we illustrated before. It is very hard for these methods to extract effective features from DR color fundus image datasets, especially for ML ones. These feature extraction methods are very effective and efficient for images with large differences in features\cite{10}. Hand-crafted features are sensitive to fundus photography quality, such as noise, equipment, and exposure intensity. The traditional classifiers such as SVM and KNN perform very well for small datasets but work poorly in large-scale datasets. Therefore, their generalization is limited. Thus, these methods are universal feature extraction methods, which may not be suitable for DR detection, which is a more challenging and specific task. In this paper, we present a novel customized CNN model (EfficientNet)\cite{11}. To diagnose the DR grades from different color fundus images automatically without making hand-crafted feature extraction or segmentation phases. A deep fine-tuned CNNs are very useful in medical image analysis and even outperform the fully trained CNNs, especially in limited training set" \cite{12,13}. We customized efficientNet with hyper parameterization to increase accuracy. 
The rest of this manuscript is organized as follows. Section \ref{sec:RW} presents all the studies currently conducted and their findings in detecting and grading DR.  Section \ref{sec:CAD} provides a detailed illustration of the proposed framework based on DL. Section 4 shows the experiments that we did and the critical reached results according to applying the proposed system on the IDRiD dataset. Finally, Section 5 presents the conclusion and future work.

\section{The Related Work}
\label{sec:RW}
  A lot of researchers pay attention to the DR diagnosing. Therefore they utilized DL in many CAD systems to detect DR. Some of these systems presented in literature depended absolutely on DL. For example, Pan et al.\cite{10} processed the color fundus images by standardizing all images to a standard size of $256\times256$. They removed noise by using a Gaussian filter. Then, they applied some data augmentation procedures such as flipping, rotation, and cropping. They utilized ResNet 18 with the regional scoring map. They employed L2 weight with the decay of $5\times10\textsuperscript{-4}$ for all trainable variables except batch-normalization (BN). They trained the model with a $10\textsuperscript{-3}$ learning rate for 200 epochs. Although the authors raised the accuracy of detecting the severity grade, it could not predict the MAs signs, which represent the mild grade. Their system was able to detect H\-EX, S\-EX, HM, and NV but could not capture MAs, contributing to the DR. 
The work that was conducted by Sahlsten et al.\cite{14} is like what was presented by Pant et al.\cite{10}, while Sahlsten et al.\cite{14} made grading of macular edema (ME). The authors started with the cropping and resizing images. They utilized deep NN (DNN) based on Inception-v3 architecture. They used instance normalization layers rather than BN layers. Although the authors predicted the severity grades accurately in DR and ME, some limitations were found in their work. The first one was concerned with the image grading reference. Grader biases affect negatively model performance. Second, DNN learns features that are unknown and ignored by experts. That is done when the network was applied on the image without determining the essential features in diagnoses such as MAs and EXs as well as their numbers, which are essential rules in grading DR.
Unlike the studies of Pant et al.\cite{10} and Sahlsten et al. \cite{14}, who used the color fundus images that were captured by fundus cameras, Rajalakshmi et al. \cite{15} used a portable fundus camera that was a smartphone-based imaging device. They utilized the EyeArt v2.1.0 application. The authors skipped the training phase. Their work's main problems are that the poor image quality according to media opacities like cataract, and they utilized small samples of the utilized datasets. 
Doshi et al. \cite{17} aimed to diagnose the DR grades without extracting features manually. Therefore, they utilized the CNN model to learn features. They made image resizing such as pant et al.\cite{10} and Sahlsten et al. \cite{13}, but they resized images to $512\times512$, convert all images to the green channel, and applied histogram equalization (HE). The authors removed noise by using Min-Max normalization. They built their model based on two consecutive convolutions (CONV) layers followed by max-pooling (MP) and dropout (DO). After that, two of three consecutive CONV layers are followed by MP and DO, then one of two consecutive CONV layers followed by MP and DO. Finally, they added two fully connected layers (FC) and output five classes due to the DR dataset class label numbers. Although the authors extracted only the green channel to save memory space, the two other red and blue channels could give more features in the learning process. Besides, they should utilize a strong filter for removing noise rather than making the normalization only.
Wang and Yang [16] were along the same lines as Doshi et al. [17]. They also built a CNN model to diagnose the grades of DR. They built two CONV layers followed by MP, then three consecutive CONV layers followed by MP. Finally, they added two successive CONV layers followed by global pooling (GP) and the FC or dense layer to output the DR classes. The authors' work recorded a better kappa score than what was achieved by Doshi et al.\cite{17}. Wang and Yang\cite{16} made some data augmentation processes such as translation, stretching, rotation, and flipping, but it is essential to reduce noise with a significant filter.
To overcome the current literature's limitations, we processed the images to enhance the contrast and remove noise. After that, we resized images to a standard size 256×256. Filtering, cropping, rotation, mirroring, and normalization were done to avoid overfitting and ensure quality. Then, we optimized the traditional effecientNet model by customizing the hyper\-parameters. After that, we evaluated the system by IDRiD dataset and the performance measures metrics.

\section{The Proposed CAD System Framework}
\label{sec:CAD}
This section presents an illustration of the proposed CAD system framework to diagnose the normal and DR grades. The system outputs five classes (normal, mild, moderate, severe NPDR, and PDR). We start our framework by feeding the Indian Diabetic Retinopathy Image Dataset (IDRiD) color fundus images to the preprocessing phase. We enhanced contrast and removed the noise. Besides, we normalized the images and resized them to a standard size of $256 \times256$. After that, we made cropping, rotation, mirroring. The normalized images are fed to the customized effecientNet model to output the healthy and DR grades, as shown in Fig. \ref{fig:CAD}. The framework consists of the following phases:
\begin{figure*}[ht]
    \centering
\includegraphics[width=\textwidth]{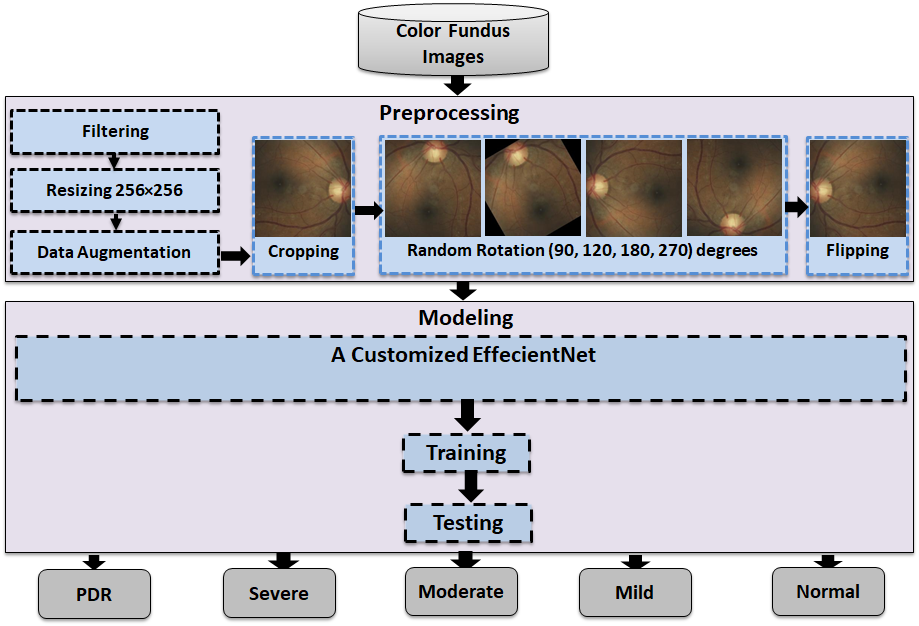}
    \caption{The Proposed CAD system Framework.}
    \label{fig:CAD}
    \end{figure*}

\subsection{The Preprocessing}
\begin{figure}[ht]
    \centering
\includegraphics[width=0.95\columnwidth]{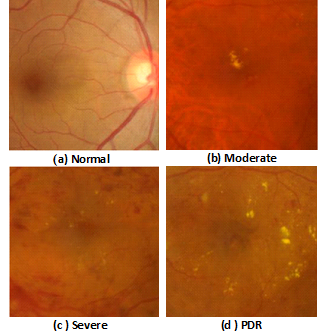}
    \caption{The example of the Cropping step.}
    \label{fig:cropping}
    \end{figure}

This phase consists of some steps. First, we make filtering by the median filter to remove noise and improve the contrast. Second, we resized the images to a standard size of $256\times256$. Third, we made some transformation processes on the images, such as cropping, random rotation, and vertical and horizontal flipping. Finally, we normalized the images by using the mean and standard deviation (STD). We make all of these preprocessing steps on the entered color fundus images to enlarge the dataset. The main advantage of learning transfer is that it can train small datasets. Moreover, we utilized the augmentation processes to avoid overfitting. Fig \ref{fig:cropping} shows the example of the cropping step in the transformation operations.

\subsection{Modeling}
EffecientNet is one of the CNN architectures. It is a scaling model that uniformly scales all dimensions of depth, width, and resolution of the network by utilizing a compound coefficient\cite{11}. Model scaling aims to increase the depth by making the network deeper, adding more convolution layers. These layers provide more detailed features from the image. Of course, the deeper model is more accurate than the models with few layers.
On the other hand, the model with more significant channels in its layer is wider than the others with few channels (R, G, and B). Moreover, the higher image resolution equals higher accuracy and more details. But the increase in the three aforementioned items results in a bigger network. So, it is very necessary to balance all the network dimensions in the scaling process.
\subsubsection{EffecientNet Model}
Constant coefficients are (d) for depth, (w) for width, and (r) for image size. If the computations are $2\textsuperscript{n}$, then we increase the network depth by $D\textsuperscript{n}$, width by $w\textsuperscript{n}$, image resolution by $R\textsuperscript{n}$. The model used compound coefficient $\phi$ to scale network width, depth, and resolution uniformly. Therefore, the model distributes the resources ($x$) on depth $d$, width $w$, and resolution $r$, as shown in equations \ref{eq:G1}$-$\ref{eq:G4}.


\begin{eqnarray}
\label{eq:G1}
d = D\textsuperscript{$\phi$}
\end{eqnarray}

\begin{eqnarray} 
\label{eq:G2}
  w = \Omega\textsuperscript{$\phi$} 
\end{eqnarray}
\begin{eqnarray} \label{eq:G3}
 r = R \textsuperscript{$\phi$} 
\end{eqnarray}

\begin{eqnarray} \label{eq:G4}
  x = D. \Omega\textsuperscript{2}. R\textsuperscript{2} \approx {2}, where D\geqslant {1}, \Omega \geqslant{1}, R \geqslant{1}
\end{eqnarray}

According to the scaling theory, the EffecientNet B0 model is scaled to obtain EffecientNet B1, B2, B3, B4, B5, B6, and B7 models. In this paper, we customized the EffecientNet Bo model by changing the values of some hyperparameters. We utilized the model to make a multi-label classification of the entered color fundus images to healthy and various DR grades.  The scaling of the utilized EffecientNet Bo model architecture is shown in Table\ref{Table:effecintb0}. We can notice the changes in the resolution r, channels c, and the number of layers L, which determines the model's depth. MB is the MobileNet model. The model consists of one CONV layer with kernel three and six layers of MB CONV layers. One layer is of MB CONV1, and five are of MB CONV6 between 3 and 5 kernels. The final layer is the CONV $1\times1$ and MP, then the dense layer or FC. The model started with r of all images $256\times256$ and ended with r equals $8\times8$ with increasing R, G, and B. We utilized Relu activation function, batch normalization (BN) 512, dropout 0.4.
\begin{table}[htb]
	\centering
\caption{\textbf{The scaling parameters of the EffecientNet B0 model.}}  \label{Table:effecintb0}
\begin{tabular}{|l|p{15mm}|p{10mm}|p{10mm}|}\hline

\textbf{Operator } &\textbf{ R} &\textbf{ c}&\textbf{ d}\\\hline 
   CONV $3\times3$&	$256\times256$&	32&	1\\\hline
   MB CONV1&	$128\times128$&	16&	1\\\hline
   MB CONV6& $128\times128$&	24&	2\\\hline
   MB CONV6&	$64\times64$&	40&	2\\\hline
   MB CONV6	&$32\times32$	&80&	3\\\hline
   MB CONV6	&$16\times16$	&112	&3\\\hline
   MB CONV6	&$16\times16$&	192&	4\\\hline
   MB CONV6&$	8\times8	$ &320&	1\\\hline
   CONV $1\times1$, MP, dense& $8\times8$&	1280&	1\\\hline
\end{tabular}
\end{table}

\section{Results}
\label{sec:Res}
\begin{figure}[ht]
    \centering
\includegraphics[width=0.95\columnwidth]{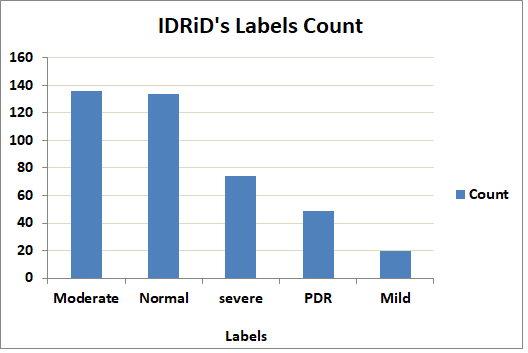}
    \caption{The distribution of the IDRiD dataset labels.}
    \label{fig:IDRiD}
    \end{figure}

\begin{figure}[ht]
    \centering
\includegraphics[width=0.95\columnwidth]{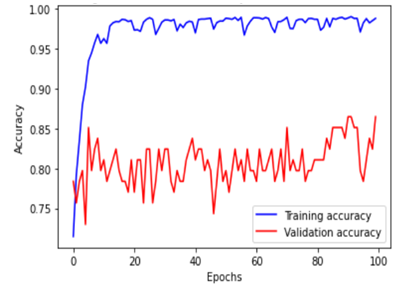}
    \caption{The training and validation Accuracy of the IDRiD dataset.}
    \label{fig:ACC}
    \end{figure}

In this paper, we utilized IDRiD\cite{12} dataset. In DR grading, the dataset contains 516 images. All of them are in JPEG. The images are split into 413 and 103 as training and testing sets, respectively. All images are $4288\times2848$ and captured by AKowa VX-10 alpha digital fundus camera with $50^o$ FOV and centered near the macula. This dataset is multi-label. It means that it contains many DR lesions. These lesions constitute the DR grade; mild, moderate, and severe. Table \ref{Table:idridDS} and Fig\ref{fig:IDRiD} show the labels of the IDRiD dataset and their counts.
\begin{table}[htb]
	\centering
\caption{\textbf{The labels and their counts according to IDRiD images.}}  \label{Table:idridDS}
\begin{tabular}{|l|p{15mm}|}\hline

\textbf{Label } &\textbf{ Count} \\\hline 
   normal&	134\\\hline
   Mild & 20\\\hline
   Moderate&	136\\\hline
   Severe & 74\\\hline
   PDR & 49\\\hline
  \end{tabular}
\end{table}
This work was implemented by using python 3.7. We ran our experiments on a core i5/2.4 GHz computer with 8 GB RAM and an NVIDIA/ (1 GB VRAM) VGA card. We measured the customized model's performance by accuracy (ACC) and Dice similarity coefficient (DSC), as shown in Eq. 3 and 4.
\begin{equation} \label{eq:ACC}
ACC=\frac{TP+TN}{TP+TN+FP+FN}
\end{equation}

\begin{equation} \label{eq:MF}
DSC =\frac{2\times TP}{2\times TP +FP+FN}
\end{equation}
where TP is a true positive, TN is a true negative, FN is a false negative, and FP is a false positive. Table\ref{Table:parameters} shows the values of hyper parameters of the customized EffecientNet B0 and the DSC results. 
From Table\ref{Table:parameters}, we can notice that we tried many optimizers such as Adam, Adagrad, Adadelta, AdaMax, ASGD, RmsProp, and Rprop. Moreover, we change the epochs from 10 to 100. From the different values, we can notice that the Adamax optimizer with 100 epochs and learning rate lr 0.002 achieved DSC to be 78.45\% in training and 65\% in the validation on the IDRiD dataset. The parameters of the AdaMax optimizer are $etas=(0.9, 0.999)$, $eps=1e-08$. The optimizer Adam comes in the third order as it achieved 56\% for DSC with the same parameters. But AdamW comes in the second-order as it achieved 63\% for DSC with $betas=(0.9,0.999)$, $eps=1e-08$, weight\_decay=0.01. RmsProp optimizer comes in the fourth-order. It achieved 62\% for DSC with 10 epochs and 0.002 lr. RmsProp is with $alpha=0.99$, and $ps=1e-08$ Similarly, we can observe that ASGD with $lambda=0.0001$, $alpha=0.75$, and $t0=1000000$ is the worst optimizer in classifying the IDRiD dataset by the EffecientNet B0 model. It is trained by 0.002 lr and 100 epochs. On the other hand, the model achieved 84\% accuracy of multi-label classification. Fig. \ref{fig:ACC} shows the ACC of the training and validation through 100 epochs.
\begin{table*}[ht]
	\centering
	\caption{The hyper-parameters values of the model on IDRiD dataset.}
	\label{Table:parameters}
	\begin{tabular}{|l|p{45mm}|p{10mm}|p{15mm}|p{15mm}|p{15mm}|p{15mm}|p{15mm}|}\hline
		\textbf{optimizer}&\textbf{optimizer parameters}&\textbf{epochs}&\textbf{$lr$}&\textbf{$T_loss$}&\textbf{$T_DSC$}&\textbf{$V_loss$}&\textbf{$V_DSC$}\\\hline
		\hline
		\textbf{Adadelta} & $\rho=0.9, \epsilon=1e-06$ & 10&	0.001&	0.74&	18&	0.68&	20 \\
		&&10&	0.001&	0.65&	24.5&	0.73&	16.2\\\hline
		\textbf{Adagrad}&$\epsilon=1e-10$&50	&0.001&	0.66&	24.12&	0.73&	15.7\\
		&&10&	0.001&	0.57&	28.2&	0.56&	25.8\\
		&&50&	0.002&	0.50&	39.4&	0.58&	26\\\hline
		\textbf{Adam}& $\beta=(0.9, 0.999), \epsilon=1e-08$&100&	0.002&	0.19&	56&	0.28&	29\\
		&&10&	0.003&	0.20&	44&	0.27&	45\\
		&&10&	0.002&	0.44&	47&	0.62&	29\\\hline
\textbf{Adamax}&$\beta=(0.9, 0.999), \epsilon=1e-08$&10&	0.002&	0.14&	65.2&	0.26&	30\\
&&50&	$le-3$&	0.56&	28&	0.58&	18\\
&&100&	0.002&	0.8&	78.45&	0.12&	65\\
&&50&	0.002&	0.48&	43&	0.50&	31\\\hline
\textbf{AdamW}&$\beta=(0.9, 0.999), \epsilon=1e-08, weight\_decay=0.01$&100&	0.002&	0.30&	63&	0.35&	36\\
&&100&	0.001&	0.72&	18&	0.69&	21\\\hline
\textbf{ASGD}&$\lambda=0.0001, \alpha=0.75, t0=10\textsuperscript{6}$&100&	0.002&	0.75&	18&	0.85&	24\\\hline
\textbf{RmsProp}&$\alpha=0.99, \epsilon=1e-08$&10&	0.001&	0.23&	28&	0.22&	61\\
&&100&	0.002&	0.42&	26&	6.7&	66\\
&&10&	0.002&	0.62&	25&	0.61&	24\\\hline
\textbf{Rprop}&$\beta=(0.5, 1.2), step\_sizes=(1e-06, 50)$&10&	0.001&	0.19&	41&	0.18&	53\\\hline
\end{tabular}
\end{table*}
\subsection{Conflicts of interest}
The authors have no conflicts of interest to declare.
\section{Conclusion}
\label{sec:Conc}
 Transfer learning is essential in a small dataset to avoid overfitting. Moreover, it reduces the training time and saves memory space. According to this fact, we utilized the pre-trained EffecientNet B0 model. Besides, depending on the multi-label classification idea, we can detect the healthy and diabetic retinopathy grades mild, moderate, severe, and PDR. We get the merits of the label correlation and dependency by using the pre-trained efficientNet B0 model.  We customized the model by changing the optimizer's hyperparameters, learning rate, and the epochs' number. The augmentation is significant, especially in the small dataset, to avoid falling into overfitting. In the future, we intend to detect other retinal diseases such as Glaucoma and age-related macular degeneration simultaneously.


\bibliographystyle{IEEEtran}
\bibliography{IEEEabrv,mybibfile}

\begin{thebibliography}{10}
\providecommand{\url}[1]{#1}
\csname url@samestyle\endcsname
\providecommand{\newblock}{\relax}
\providecommand{\bibinfo}[2]{#2}
\providecommand{\BIBentrySTDinterwordspacing}{\spaceskip=0pt\relax}
\providecommand{\BIBentryALTinterwordstretchfactor}{4}
\providecommand{\BIBentryALTinterwordspacing}{\spaceskip=\fontdimen2\font plus
\BIBentryALTinterwordstretchfactor\fontdimen3\font minus
  \fontdimen4\font\relax}
\providecommand{\BIBforeignlanguage}[2]{{%
\expandafter\ifx\csname l@#1\endcsname\relax
\typeout{** WARNING: IEEEtran.bst: No hyphenation pattern has been}%
\typeout{** loaded for the language `#1'. Using the pattern for}%
\typeout{** the default language instead.}%
\else
\language=\csname l@#1\endcsname
\fi
#2}}
\providecommand{\BIBdecl}{\relax}
\BIBdecl

\bibitem{1}
``{At a glance: Diabetic Retinopathy},''
  \url{https://www.nei.nih.gov/learn-about-eye-health/eye-conditions-and-diseases/diabetic-retinopathy},
  [Online; accessed 1-oct-2020].

\bibitem{2}
``{Webmd},'' \url{https://www.webmd.com/diabetes/diabetic-retinopathy},
  [Online; accessed 2-oct-2020].

\bibitem{3}
W.~L. Alyoubi, W.~M. Shalash, and M.~F. Abulkhair, ``Diabetic retinopathy
  detection through deep learning techniques: A review,'' \emph{Informatics in
  Medicine Unlocked}, p. 100377, 2020.

\bibitem{4}
S.~Dhingra and P.~T. Khaw, ``The moorfields safer surgery system,''
  \emph{Middle East African journal of ophthalmology}, vol.~16, no.~3, p. 112,
  2009.

\bibitem{5}
N.~Dey, S.~Borra, A.~S. Ashour, and F.~Shi, \emph{Machine Learning in
  Bio-Signal Analysis and Diagnostic Imaging}.\hskip 1em plus 0.5em minus
  0.4em\relax Academic Press, 2018.

\bibitem{6}
J.~Gao, C.~Leung, and C.~Miao, ``Diabetic retinopathy classification using an
  efficient convolutional neural network,'' in \emph{2019 IEEE International
  Conference on Agents (ICA)}.\hskip 1em plus 0.5em minus 0.4em\relax IEEE,
  2019, pp. 80--85.

\bibitem{7}
B.~Tymchenko, P.~Marchenko, and D.~Spodarets, ``Deep learning approach to
  diabetic retinopathy detection,'' \emph{arXiv preprint arXiv:2003.02261},
  2020.

\bibitem{8}
D.~Kirange, J.~Chaudhari, K.~Rane, K.~Bhagat, and N.~Chaudhri, ``Diabetic
  retinopathy detection and grading using machine learning,'' \emph{Int. J. of
  Adv. Trends in Comput. Sci. \& Engine}, vol.~8, no.~6, pp. 3570--3576, 2019.

\bibitem{9}
K.~Xu, D.~Feng, and H.~Mi, ``Deep convolutional neural network-based early
  automated detection of diabetic retinopathy using fundus image,''
  \emph{Molecules}, vol.~22, no.~12, p. 2054, 2017.

\bibitem{10}
P.~Junjun, Y.~Zhifan, S.~Dong, and Q.~Hong, ``Diabetic retinopathy detection
  based on deep convolutional neural networks for localization of
  discriminative regions,'' in \emph{2018 International Conference on Virtual
  Reality and Visualization (ICVRV)}.\hskip 1em plus 0.5em minus 0.4em\relax
  IEEE, 2018, pp. 46--52.

\bibitem{11}
M.~Tan and Q.~V. Le, ``Efficientnet: Rethinking model scaling for convolutional
  neural networks,'' \emph{arXiv preprint arXiv:1905.11946}, 2019.

\bibitem{12}
P.~Porwal, S.~Pachade, R.~Kamble, M.~Kokare, G.~Deshmukh, V.~Sahasrabuddhe,
  F.~Meriaudeau, G.~Quellec, T.~MacGillivray, L.~Giancardo \emph{et~al.},
  ``Diabetic retinopathy: Segmentation and grading challenge workshop,'' in
  \emph{Proceedings of the IEEE International Symposium on Biomedical Imaging
  (ISBI’18)}, 2018.

\bibitem{13}
K.~Simonyan and A.~Zisserman, ``Very deep convolutional networks for
  large-scale image recognition,'' \emph{arXiv preprint arXiv:1409.1556}, 2014.

\bibitem{14}
J.~Sahlsten, J.~Jaskari, J.~Kivinen, L.~Turunen, E.~Jaanio, K.~Hietala, and
  K.~Kaski, ``Deep learning fundus image analysis for diabetic retinopathy and
  macular edema grading,'' \emph{Scientific reports}, vol.~9, no.~1, pp. 1--11,
  2019.

\bibitem{15}
R.~Rajalakshmi, R.~Subashini, R.~M. Anjana, and V.~Mohan, ``Automated diabetic
  retinopathy detection in smartphone-based fundus photography using artificial
  intelligence,'' \emph{Eye}, vol.~32, no.~6, pp. 1138--1144, 2018.

\bibitem{17}
D.~Doshi, A.~Shenoy, D.~Sidhpura, and P.~Gharpure, ``Diabetic retinopathy
  detection using deep convolutional neural networks,'' in \emph{2016
  International Conference on Computing, Analytics and Security Trends
  (CAST)}.\hskip 1em plus 0.5em minus 0.4em\relax IEEE, 2016, pp. 261--266.

\bibitem{16}
Z.~Wang and J.~Yang, ``Diabetic retinopathy detection via deep convolutional
  networks for discriminative localization and visual explanation,''
  \emph{arXiv preprint arXiv:1703.10757}, 2017.

\end{thebibliography}




%



\end{document}